\documentclass[a4paper,fleqn,usenatbib]{mnras}

\usepackage{newtxtext,newtxmath}

\usepackage[T1]{fontenc}
\usepackage{ae,aecompl}

\usepackage{graphicx}	
\usepackage{amsmath}	
\usepackage{amssymb}	

\title[Sextans Star Formation History]{The star formation history of the Sextans dwarf spheroidal galaxy: a true fossil of the pre-reionization era}

\author[M. Bettinelli et al.]{
M. Bettinelli,$^{1,2,3}$\thanks{E-mail: mbettine@iac.es}
S.~L. Hidalgo$^{1,2}$,
S.~Cassisi$^{4,1}$,
A.~Aparicio$^{2,1}$ and
G.~Piotto$^{3,5}$
\\
$^{1}$Instituto de Astrof\`isica de Canarias, V\`{i}a L\`{a}ctea S/N, E-38200 La Laguna, Tenerife, Spain\\
$^{2}$Department of Astrophysics, University of La Laguna, E-38200 La Laguna, Tenerife, Canary Islands, Spain\\
$^{3}$Dipartimento di Fisica e Astronomia ``Galileo Galilei'', Universit\`{a} degli Studi di Padova,  Vicolo dell'Osservatorio 3, I-35122 Padova, Italy\\
$^{4}$INAF-Osservatorio Astronomico d'Abruzzo, Via M. Maggini, sn. 64100 Teramo, Italy\\
$^{5}$INAF-Osservatorio Astronomico di Padova, Vicolo dell'Osservatorio 5, I-35122 Padova, Italy\\
}

\date{Accepted 2018 January 19. Received 2018 January 11; in original form 2017 October 2}
\pubyear{2018}

\begin{document}
\label{firstpage}
\pagerange{\pageref{firstpage}--\pageref{lastpage}}
\maketitle

\begin{abstract}
We present the star formation history (SFH) of the Sextans dwarf spheroidal galaxy based on deep archive $B$,$I$ photometry taken with Suprime-Cam at Subaru telescope focusing our analysis on the inner region of the galaxy, fully located within the core radius. Within the errors of our SFH we have not detected any metallicity gradient along the considered radial distance interval. As a main result of this work we can state that the Sextans dwarf spheroidal stopped forming stars less than $\sim1.3$ Gyr after Big Bang in correspondance to the end of the reionization epoch. We have been able to constrain the duration of the main burst of star formation to $\sim0.6$ Gyr.
From the calculation of the mechanical luminosity released from supernovae (SNe) during the brief episode of star formation, there are strong indications that SNe could have played an important role in the fate of Sextans, by removing almost completely the gas component, so preventing a prolonged star formation.
\end{abstract}




\section{Introduction}
Dwarf galaxies are the more common and diffused type of galaxies in the Universe. Precisely for this reason, their deep understanding could strongly influence the theories of galaxy formation and evolution. In the $\Lambda$ cold dark matter ($\Lambda$CDM) scenario dwarf galaxies are building blocks of larger systems (\citealt{1985Natur.313...72B}, \citealt{1997ApJ...490..493N} ). Seen in this light, they are the survivors of early merging processes that led to the formation of bigger galaxies such as the Milky Way (MW). 
Thanks to their proximity, MW satellite galaxies, can be resolved into stars. From deep colour-magnitude diagrams (CMDs) reaching the oldest main sequence (MS) turn-off, the star formation history (SFH) can be derived displaying all the stars born during the lifetime of the system \citep{2016MmSAI..87..350H}.

This work is part of a larger project aimed to analyze and constrain the early star formation activity in a sample of MW dwarf spheroidal (dSph) galaxies. The main goal is to discern whether these galaxies are true fossils of the Reionization era \citep{2005ApJ...629..259R}. 
There is evidence that these objects present a very old stellar population produced mostly in a single episode of star formation at early times (\citealt{2004ApJ...610L..89G}, \citealt{2014ApJ...789..148W}). The ultimate objective of our analysis is to find the imprints in the SFH of the physical phenomena that have determined the complete quenching of star formation, such as the global cosmic reionization or local effects like SNe feedback or tidal interactions.
The first case analyzed in the following work is the one of Sextans.

The Sextans dwarf spheroidal was discovered relatively recently by \citet{1990MNRAS.244P..16I} during a quasars survey. It has not been possible to detect it simply by visual inspection, as it was done for the other seven classic Sculptor-like dwarfs, due to its very low central surface brightness, estimated to be $18.2\pm0.5$ mag/arcmin$^{2}$ \citep{1995MNRAS.277.1354I}.
Since its discovery many photometric surveys have been performed in order to investigate the stellar populations properties of the system. In the first data available of \citet{1990MNRAS.244P..16I} a red horizontal branch and a well defined asymptotic giant branch were already clearly present. 
\citet{1991AJ....101..892M} data showed the MS turn-off, a defined red giant branch and the presence of blue stragglers (BS).
In the the analysis performed by \citet{2001MNRAS.327L..15B} (see also \citet{2002ASPC..274..421P}) the authors discovered the bimodal distribution in color of the red giant branch (RGB) stars and a double RGB-bump, unequivocally sign of two distict stellar components, a main one with $[Fe/H]\sim-1.8$ and a minor one with $[Fe/H]\sim-2.5$.
\citet{2003AJ....126.2840L} presented deep wide-field BVI photometry obtained at the $3.6 \,{\rm m}$ Canada-France-Hawaii Telescope (CFHT); later, \citet{2009ApJ...703..692L} derived the star formation history (SFH) making use of VI photometry. Interestingly they found, as \citet{2002ASPC..274..421P}, that in the central regions the metallicity increases up to $[Fe/H]\sim-1.6$, while in the more external part of the investigated field metallicity decreases down to $[Fe/H]\sim-1.8$. 
The spectroscopic determinations presented in literature are systematically more metal poor: \citet{2011ApJ...727...79K} measured a median $[Fe/H]=-2.00$. \citet{2003AJ....126.2840L} measured a similar value making use of the $(V-I)$ colour relation defined in \citet{1993ApJ...408..409L}, finding a mean metallicity of the RGB of $ [Fe/H]\sim-2.1\pm0.1$, which would agree with the correlation between $[Fe/H]$ vs. $M_{V}$ followed by other dwarf galaxies (e.g. \citet{2003AJ....125.1926G}, \citet{2012AJ....144....4M}).

From the structural point of view, \citet{2016MNRAS.460...30R} published an extensive photometric study up to $83$ arcmin from Sextans' centre. The authors do not find any trace of tidal disruption but have been able to detect some overdensities equally distributed around the centre.
\citet{2017MNRAS.tmp...97O} detected the existence of an age gradient with younger stars more centrally concetrated than older ones, thus suggesting that it is unlikely that the inner region of Sextans could have been perturbed by strong tidal disruption. 

The most relevant physical parameters of Sextans are summarized in Table~\ref{tab:param}.

In this paper we present the analysis of the SFH of the Sextans dwarf spheroidal based on deep wide-field, ground-based photometry. 
The paper is organized as follows: in \S~\ref{sec:observations}, the observations, data reduction, the derivation of the photometry and the photometric calibration are presented. In \S~\ref{sec:cmd} the colour-magnitude diagram (CMD) of Sextans is presented.
In \S~\ref{sec:dersfh} presents the procedure of the SFH derivation; in \S~\ref{sec:confining} it is confined the first event of star formation. In \S~\ref{sec:radial} we performed a radial study of the SFH. In \S~\ref{sec:disc} presents a discussion about the achieved results. Finally, in \S~\ref{sec:conc} there are the summary and conclusions.

\begin{table*}
\begin{minipage}{118mm}
\caption{Sextans parameters}
\label{tab:param}
\begin{tabular}{@{}lcr}
\hline
Quantity & Value & References\footnote{(1) \citet{1990MNRAS.244P..16I};  (2) \citet{1998ARA&A..36..435M}; (3) \citet{1994MNRAS.269..957H}; (4) \citet{2016MNRAS.460...30R};  (5) \citet{1995MNRAS.277.1354I}; (6) \citet{2009MNRAS.394L.102L}.}\\
\hline
RA, $\alpha$ (J2000.0) &  $10^{\rm h} 13^{\rm m} 03^{\rm s}.0$ & (1)\\ 
Dec, $\delta$ (J2000.0) & -1$^{\circ}$ 36$^\prime$ 52.0$^{\prime\prime}$ & (1)\\ 
Galactic longitude, $l$ ($^\circ$) & 243.5 & (1)\\ 
Galactic latitude, $b$ ($^\circ$) & 42.3 & (1)\\ 
Galactocentric distance (kpc) & 86$\pm$5 & (2)\\ 
Heliocentric velocity (km s$^{-1}$) \hspace{20pt}& 224.4$\pm$1.6 & (3)\\ 
Ellipticity, $e$ & $0.29\pm 0.03$ & (4)\\ 
Position angle ($^\circ$) & 56.7$\pm$2.8 & (4) \\ 
Core radius ($^\prime$) & 26.8$\pm$1.2 & (4)\\ 
Tidal radius ($^\prime$) & 83.2$\pm$7.1 & (4)\\ 
Luminosity, $L_V$ ($L_\odot$) & ($4.1\pm1.9$)$\times 10^5$ & (5) \\ 
Absolute magnitude, $M_V$ & $-9.2\pm 0.5$ & (5)\\ 
Total mass, ($M_\odot$) & (4.0$\pm0.6$)$\times 10^7$ & (6) \\ 
Mass to light ratio, $M_\odot/L_\odot$ & $91\pm49$ & (6) \\ 
\hline
\end{tabular}
\end{minipage}
\end{table*}


\section{Observations and Data Reduction}
\label{sec:observations}
The deep wide-field photometry we present in this work is a stack of many archive observations in $B$ and $I$ filters, (see Table~\ref{tab:obs} for details), covering approximately an area of $34 \times 27$ arcmin centered on the galaxy, taken with the Subaru Prime Focus Camera, Suprime-Cam \citep{2002PASJ...54..833M} in years $2003$ and $2005$. We have selected the data making use of the Subaru-Mitaka-Okayama-Kiso Archive System (SMOKA) \citep{2002ASPC..281..298B}. Reduction, that means overscan, bias substraction, flat fielding, distortion and atmospheric dispersion corrections (see \citet{2002PASJ...54..833M}) and sky substraction, was performed using the SDFRED1 software (\citealt{2002AJ....123...66Y}; \citealt{2004ApJ...611..660O}).
The total exposure time is $2800$ s in $B$ filter and $8936$ s in $I$ filter. The pixel scale is equal to $0.2''/px$.


\begin{table*}
\centering
\caption{Observations}
\label{tab:obs}
\begin{tabular}{cccccc} 
\hline
Filter & UT & PI & Exposure(s) & FWHM ('')\\
\hline
$B$& $2005/05/04$ & Kleyna J.T. & $500\times5$ & $0.6''$\\
$B$& $2005/05/04$ & Kleyna J.T. & $60\times5$ & $0.7''$\\
$I$& $2005/02/07$ & Arimoto N. & $270\times7$ & $1.0''$\\
$I$& $2003/04/01$ & Yasuda N. & $30\times5$ & $1.0''$\\
$I$& $2003/04/03$ & Ikuta C. & $240\times5$ & $0.6''$\\
$I$& $2005/05/04$ & Kleyna J.T. & $200\times15$ & $0.6''$\\
$I$& $2005/05/04$ & Kleyna J.T. & $60\times5$ & $0.6''$\\
$I$& $2005/12/31$ & Arimoto N.& $235\times10$ & $0.9''$\\
$I$& $2005/12/31$ & Arimoto N. & $30\times3$ & $0.9''$\\
\hline
\end{tabular}
\end{table*}

The photometry on stacked images was performed making use of the {\sc daophot/allstar} suite of programs \citep{1990ASPC....8..289S}. We used a Moffatian of parameter $\beta=1.5$ and radius of $R_{PSF}=15$ pixels as point-spread function (PSF). 
From the output of {\sc allstar}, only objects with $\sigma\leqslant0.2$ and $-0.5\leqslant$SHARP$\leqslant 0.5$, were taken into consideration, in order to clean up bad measured stars and galaxies.
Once obtained the two catalogues in $B$ and $I$ filters, we performed the match using the package {\sc daomaster} \citep{1993spct.conf..291S}.
From this match, the final total catalogue in $B$ and $I$ filters counts $\sim  86000$ stars.
The data have been calibrated making use of Stetson photometric standard field for Sextans (\citealt{2000PASP..112..925S}; \citealt{2005PASP..117..563S}), centered on the galaxy, with an extension of $36.1 \times 36.0$ arcmin.
We calibrated the entire photometric catalogue derived from the stacked images, applying the following system of two equations:

\begin{align}
\label{eqn:calib2}
\begin{split}
B-b &= c_{b}\times(B-I)+z_{b}
\\
I-i &= c_{i}\times(B-I)+z_{i}
\end{split}
\end{align}

where $B$, $I$ are the standard magnitudes, $b$, $i$ the instrumental magnitudes ,$c_{b}$,  $c_{i}$  the colour therms and $z_{b}$, $z_{i}$, the photometric zero points. 
All the coefficients relative to the two derived photometric transformations, with the corresponding errors, are listed in Table~\ref{tab:calib}.

\begin{table}
\centering
\caption{Parameters used for calibrating the photometry with the associated errors}
\label{tab:calib}
\begin{tabular}{cc} 
\hline
Parameter & Values\\
\hline
$z_{b}$ & $27.023\pm0.005$\\
$c_{b}$ & $0.064\pm0.002$\\
\hline
$z_{i}$ & $26.865\pm0.014$\\
$c_{i}$ & $0.056\pm0.007$\\
\hline
\end{tabular}
\end{table}

\begin{figure}
\includegraphics[width=\columnwidth]{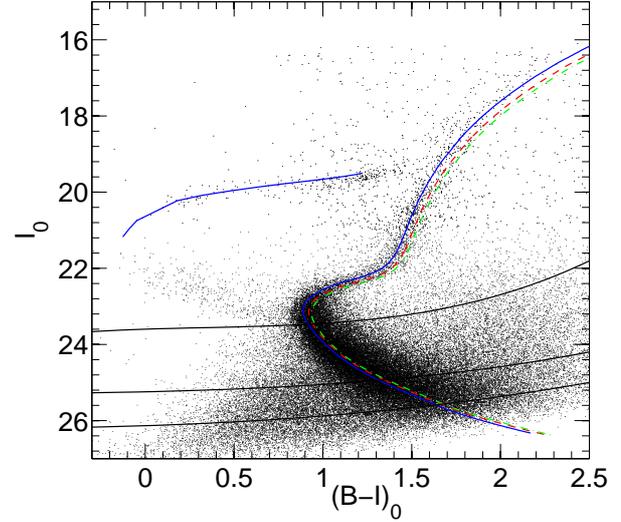}
   \caption{Observed CMD of Sextans dSph spheroidal. Three isochrones from the BaSTI stellar evolutionary library have been superimposed on the CMD: $Z=0.0003$, $13.5$ Gyr (solid blue line in the electronic version), $Z=0.0006$, $13.5$ Gyr (dash-dotted green line in the electronic version), $Z=0.0005$, $13.5$ Gyr (dashed red line in the electronic version). Red dashed line corresponds to the mean recovered metallicity for the system $[Fe/H]=-1.6$. The core helium-burning locus is given only for the $Z=0.0003$, $13.5$ Gyr isochrone (solid blue line in the electronic version). Completeness levels are over plotted as black lines: in order of increasing $I_{0}$ magnitude these are $90\%$, $75\%$ and $50\%$}
\label{fig:cmd0}
\end{figure}

\section{The colour-magnitude diagram}
\label{sec:cmd}
The CMD derived from our photometry is extended well below the oldest MS turn-off, for about $3$ magnitudes. This means that we are able to extract information till the very early epochs of Sextans and to constrain its first star formation episode. Fig.~\ref{fig:cmd0} shows the obtained CMD, corrected by reddening ($A_{B}=0.171$, $A_{I}=0.071$ (\citealt{1998ApJ...500..525S}; \citealt{2011ApJ...737..103S})), where we over plotted three isochrones from the BaSTI stellar evolution library \citep{2004ApJ...612..168P} using a distance modulus of $(m-M)_{0}=19.67$ \citep{1995AJ....110.2166M}. 
The blue and green isochrones, which correspond respectively to a metallicity of $[Fe/H]=-1.8$, $[Fe/H]=-1.5$ and an age of $13.5$ Gyr, fit well the MS, the subgiant branch (SGB) and the red-giant branch (RGB).
Nevertheless, from the broadening of the RGB locus \citep{2001AJ....122.3092H} there is some hint that there could exist a metallicity spread between $[Fe/H]=-1.8$ and $-1.5$. Although, one should note that this color broadening of the RGB could be also compatible with an age spread at fixed metallicity as suggested by \citet{2017MNRAS.tmp...97O}.
It is also present in the derived CMD a clear and extended blue and red horizontal branch (HB) and a quite populous region occupied by genuine blue stragglers \citep{2017MNRAS.tmp...97O}, that means primordial binary stars that evolved in mass-transfer BS \citep{2007A&A...468..973M}.
The field contamination level in the central region is quite low and it has been estimated to be $\sim 2 \%$ by \citet{2017MNRAS.tmp...97O}, so it does not significantly affect our results. Errors and completeness of the photometry are discussed in \S~\ref{sec:case}. 

\section{Derivation of the SFH}
\label{sec:dersfh}
For solving the SFH of Sextans we have followed the prescription outlined in \citet{2009IAUS..258..245H} and \citet{2011ApJ...730...14H}.
This method allows to derive the star formation rate as a function of time and the age-metallicity relation.
The three main codes used are: (1) IAC-Star \citep{2004AJ....128.1465A}, that computes synthetic CMDs (sCMDs); (2) IAC-pop \citep{2009AJ....138..558A}, the core algorithm for the calculation of the SFH solutions; (3) MinnIAC \citep{2011ApJ...730...14H}, a suite of routines that manages the process of sampling the parameter space, creating input data and averaging solutions.
We define the SFH as a distribution function $\psi(t,z)dtdz$ which is the usual SFR but with the dependence both on time and metallicity. 

\subsection{The case of Sextans}
\label{sec:case}
Firstly, making use of the code {\sc IAC}-star \citep{2004AJ....128.1465A} and the BaSTI \citep{2004ApJ...612..168P} stellar evolutionary library, we computed a sCMD to be used for the analysis of the Sextans stellar properties, with $5\times10^{6}$ stars, characterized by a constant star formation rate (SFR) between $0$ and $13.5$ Gyr and an uniform distribution of metallicity of $0.0001\leqslant$ Z $\leqslant 0.002$ for all ages. This metallicity range has been fixed acordingly to the metallicity distribution function (MDF) by \citet{2011ApJ...727...78K}. 
As bolometric corrections we adopted \citet{2004astro.ph..5087C}.
We used as initial mass function (IMF) the one by \citet{2002Sci...295...82K}.
For the binary star distribution $\beta(f,q_{min})$ we adopted a fraction of binary stars of $f=0.3$ and a flat distribution of the secondary to primary stellar mass ratio, $q$, with minimum $q_{min}=0.5$.

As explained in deep details in \citep{1995AJ....110.2105A}, the simulation of observational effects is a key step for the interpretation of real data. 
In particular, to obtain a realistic model CMD to be compared with the real one, we need to simulate the observational effects that affect the real data. 
In order to estimate the completeness and uncertainties of our photometry we followed the standard technique of injecting a list of artificial stars in each stacked image and then re-deriving the photometry in the same way as done for real stars (see \citet{2011ApJ...730...14H}).
We injected $5\times10^{6}$ stars in each image along an uniform grid, with a separation of at least $2 \times R_{PSF}+1$ pixels between the centroids of the artificial stars.
The stars have been chosen in order to cover the full range of luminosity and colour of the oCMD, $0\leqslant (B-I) \leqslant 2.5$ and a magnitude range of $16 \leqslant I \leqslant 26$; in particular we populated more the MS region $ 21.5 \leqslant I \leqslant 26$, with $3.5\times10^{6}$ stars, since it is the more relevant region for the derivation of the SFH, then the red-giant-branch (RGB) region $18 \leqslant I \leqslant 21.5$, with $1\times10^{6}$ stars, and finally, the upper part of the colour-magnitude diagram in the magnitude range of $16 \leqslant I \leqslant 18$, with $0.5\times10^{6}$ stars.
In Fig.~\ref{fig:cmd0}, we have over-plotted on the CMD of Sextans the completeness levels at $50\%$, $75\%$ and $90\%$. The completeness has been calculated as the ratio of the number of artificial stars recovered to the number of all injected stars in each colour and magnitude interval.  
For each artificial star we have recorded the injected magnitude $m_{inj}$, the recovered magnitude $m_{rec}$ and the position on the frame.
We use the difference between the injected and recovered magnitudes  $(m_{inj}-m_{rec})$ and the position of the synthetic star in the image to simulate the observational effects on the observed CMD. See \citet{2011ApJ...730...14H} for a full description of the procedure.

\begin{figure}

\includegraphics[width=\columnwidth]{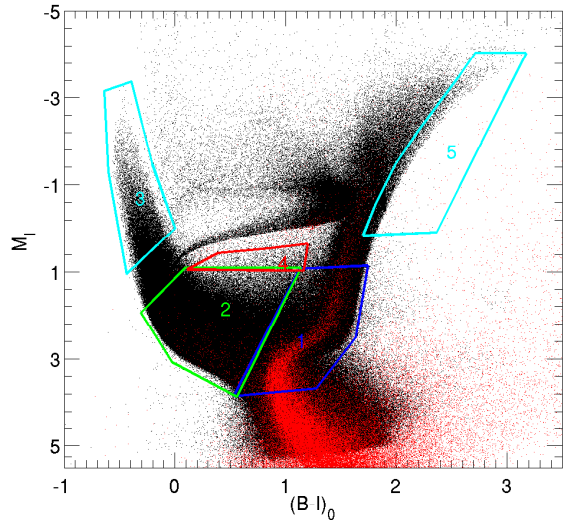}
   \caption{Synthetic CMD with the simulated observational effects (black) with overplotted the observed CMD (red). \textit{Bundles} are the regions numbered from $1$ to $5$, see text for details.}
\label{fig:boxes}
\end{figure}

In Fig.~\ref{fig:boxes} we show as black dots the recovered stars with observational effects simulation and overplotted on them as red dots the observed stars.
After the observational effects simulation, the distribution of stars in the oCMD is compared with the distribution of stars of each simple stellar population (SSPs) in the sCMD. The SSPs are selected by defining age and metallicity bins for the synthetic stars. For the age we have chosen intervals of $2$ Gyr for the first $8$ Gyr and a finer one, of $0.5$ Gyr, for the last interval between $8$ Gyr and $13.5$ Gyr, since it is necessary higher resolution for well characterize the first event of SFH.  
The metallicity bins used are: $(0.01,\,\,\, 0.1,\,\,\,  0.3,\,\,\,  0.5,\,\,\,  1,\,\,\,  2) \times 10^{-3}$. The higher the metallicities, the wider the intervals.
Both CMDs are sampled by using \textit{bundles}, macroregions which sample main features of the CMDs. Each of these bundles has a weight on the solution, given by the number of boxes defined in them. The larger the number of boxes, the larger the weight in the SFH.

We have defined five bundles, subdivided in boxes with varying size according to the region sampled, (see Table~\ref{tab:boxbundle} and Fig.~\ref{fig:boxes}).
In bundle $1$ a finer grid has been used than in the other bundles, for a total of circa $700$ boxes, since it is the region of the CMD of higher interest for the derivation of the SFH. This bundle, in fact, is the more important one for the SFH derivation because it samples the region of the CMD that is more affected by the age diffences, in the sense that it corresponds to the more reliable 'clock' provided by the CMD of any stellar populations, that is the MS turn-off.
Moreover, for the region of the MS stellar evolution models are accurate and well established.
Bundle $2$ samples the young MS, even though in the case of Sextans these stars are recognized to be genuine blue-stragglers stars \citep{2017MNRAS.tmp...97O}. The grid has a larger binning since the density of stars is lower; in this case the boxes are about $150$.
Bundle $3$ refers to the region of the MS occupied by the youngest stars, the so called blue plume; in our observed CMD there are no stars in this region, but we sample it in order to constrain the resulting SFH; the boxes in this case are $20$.
Bundle $4$ contains a portion of SGB stars (about $70$ boxes) and finally, bundle $5$ is an almost empty bundle with the largest box size; it constrains the maximum allowed metallicity. 

In order to minimize the dependence of the solution on the CMDs sampling parameters, we obtained 24 solutions varying the CMD binning within each bundle and the SSPs sampling. To do so, we have shifted by a $30\%$ the age and metallicity in a total of $12$ combinations of SSPs, each one sampled with two slightly different combinations of boxes distribution in the bundles.

\begin{figure*}

\includegraphics[width=15cm]{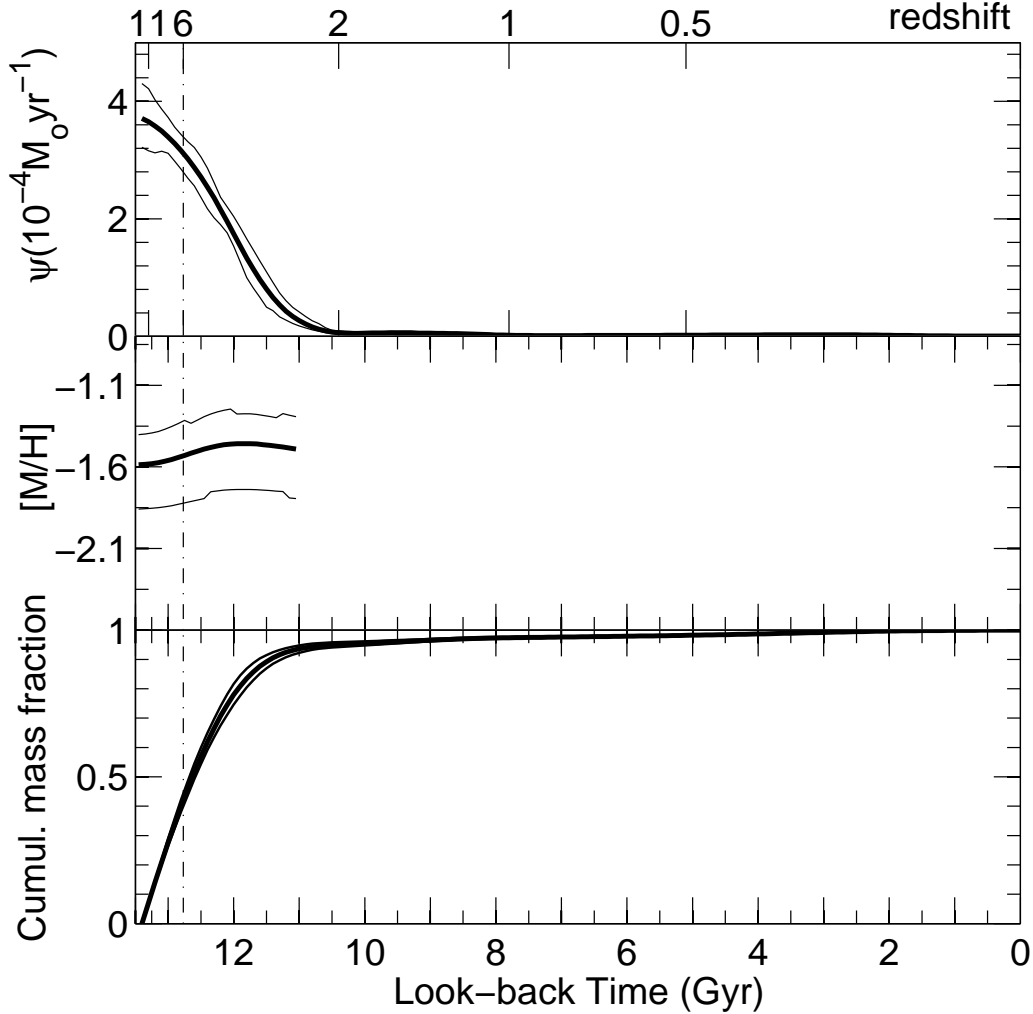}
   \caption{Results of the Sextans SFH. Top panel: SFH as a function of time ($\psi(t)$). Middle panel: metallicity of the system as a function of the time. Lower panel: cumulative mass fraction as a function of the time. Uncertainties have been drawn as this lines. }
\label{fig:sfhall}
\end{figure*}

\begin{figure*}

\includegraphics[width=15cm]{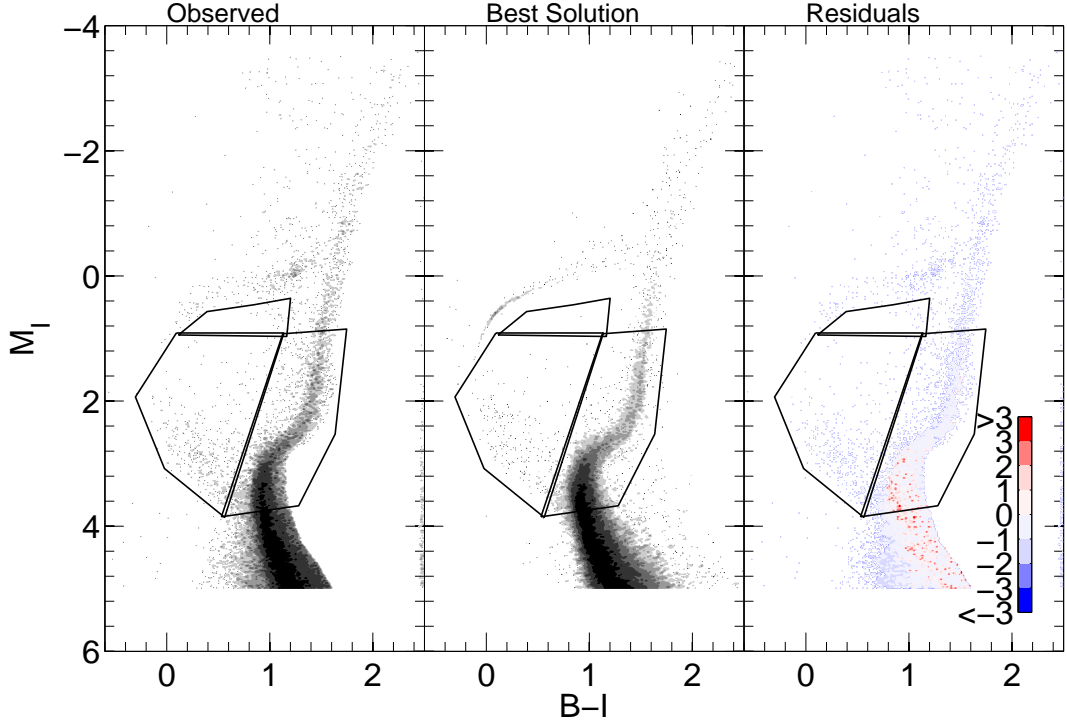}
   \caption{Hess diagrams relative to the observed CMD (left panel), best solution CMD (middle panel) and residuals CMD (right panel). The residuals are in units of Poisson uncertainties. Residuals from $\sigma=0$ to $\sigma=3$ refer to the case in which the model predicts more stars in respect to the oCMD. Residuals from $\sigma=0$ to $\sigma=-3$ refer to the case in which the model predicts less stars in respect to the oCMD. Gray levels show the density of stars. A factor of $2$ in density exists between each two successive gray levels. The single dots are shown where the density is less than $2$ stars per $(0.02)^2$ mag. The boxes show the areas of the CMD used for the derivation of the SFH.}
\label{fig:comp}
\end{figure*}

To limit the effects on the solution due to uncertainties related to the distance modulus, photometric calibration and reddening we have shifted the oCMD 25 times along a regular grid with nodes in colour $\Delta(B-I)=[-0.1,-0.05,\,\, 0,\,\, 0.05,\,\, 0.1]$ and magnitude $\Delta I=[-0.2, -0.1,\,\, 0,\,\, 0.1,\,\, 0.2]$. 
For each node we have calculated the $24$ solutions described above, obtaining a total of $600$ solutions. 
For each node a mean solution $\overline{\psi}$ and its $\overline{\chi^{2}_{\nu}}$ is calculated. In this way we have obtained $25$ different $\overline{\chi^{2}_{\nu}}$, the one with the minimum value indicates the best solution. 
The best solution we have found is the one with a shift of $+0.1$ mag in I filter and no shift in colour.
So that we are able to minimize the effects of the aforementioned external parameters.
In Fig.~\ref{fig:sfhall} are plotted the obtained results.
In the upper panel it is shown the star formation rate as a function of time, $\psi(t)$, in the middle panel the age-metallicity relation $Z(t)$ and in the lower panel the cumulative mass function. 
The resolution of our results shows that Sextans has experienced a single event of star formation limited to the first $\sim 2$ Gyrs after Big Bang.
For clarity we have tabulated the percentage of stellar mass produced in Table~\ref{tab:cumulativemass}; it is evident that Sextans has finalized the bulk of its star formation before $\sim 11.9$ Gyr ago with the peak reached at $\sim13$ Gyr ago.
In Fig.~\ref{fig:comp} the Hess diagrams of the oCMD (left panel), the best solution CMD (middle panel) and the residuals (right panel) are plotted. Note here, that we have excluded foreground contamination points in the right part of the CMD. From the residuals Hess diagram, with the values expressed in units of Poisson error, the good agreement between the observed and simulated CMD in almost all the evolutionary phases is noticeable. We note that the numerical code used for computing the synthetic CMD does not account for a dispersion of stellar masses along the core He-burning sequence as due to the occurrence of
stochastic mass loss efficiency in the previous RGB stage. This prevents the possibility to properly reproduce the color distribution of stars along the HB in the observed CMD. Nevertheless, we wish to note that this evolutionary sequence is not accounted for in the analysis of the SFH and, hence this limitation has no impact at all on the results presented in this paper. It is interesting to note that objects at faint magnitudes to the blue side of the MS in the observed CMD are largely absent in 'best solution' CMD; these are probably unresolved faint galaxies.


\begin{table}
\centering
\caption{Box sizes in each bundle that sample the observed CMD}
\label{tab:boxbundle}
\begin{tabular}{ccc} 
\hline
Bundle \# & $\Delta col$ & $\Delta mag$\\
\hline
$1$ & $0.025$ & $0.125$\\
$2$ & $0.1$ & $0.205$\\
$3$ & $0.1$ & $0.23$\\
$4$ & $0.1$ & $0.23$\\
$5$ & $1.5$ & $0.67$\\
\hline
\end{tabular}
\end{table}



\begin{table}
\centering
\caption{Mass percentiles formed in Sextans as a function of redshift and look-back time.}
\label{tab:cumulativemass}
\begin{tabular}{ccc} 
\hline
Mass \% & $Look-back\,Time (Gyr)$ & $Redshift$ \\
\hline
$10$ & $13.26$ & $10.1$\\
$20$ & $13.11$ & $8.3$\\
$30$ & $12.95$ & $7$\\
$40$ & $12.80$ & $6.1$\\
$50$ & $12.64$ & $5.4$\\
$60$ & $12.43$ & $4.7$\\
$70$ & $12.22$ & $4.15$\\
$80$ & $11.92$ & $3.55$\\
$90$ & $11.45$ & $2.9$\\ 

\hline
\end{tabular}
\end{table}
\section{Confining the first event of star formation}
\label{sec:confining}
Uncertainties deriving from observational data and from the SFH computational procedure affect the shape of the final SFH, in the sense that there is a loss of age resolution (\citealt{2009AJ....138..558A}, \citealt{2011ApJ...730...14H}, \citealt{2016ApJ...823....9A}). This effect translates into an amplification of the duration of the main event of SFH.
In order to confine the duration of the first stellar burst in Sextans we have calculated the SFH of a number of mock stellar populations with an increasing age width all starting $13.5$ Gyr ago. The duration of each burst, in terms of Full Width at Half Maximum ($FWHM_{in}$), has been chosen to be: $0.25$, $0.5$, $0.75$, $1$, $1.5$, $2$ Gyr.
Metallicity has been fixed to $[Fe/H]=-1.6$, which is the mean metallicity recovered for the total SFH of Sextans.

\begin{figure}

\includegraphics[width=\columnwidth]{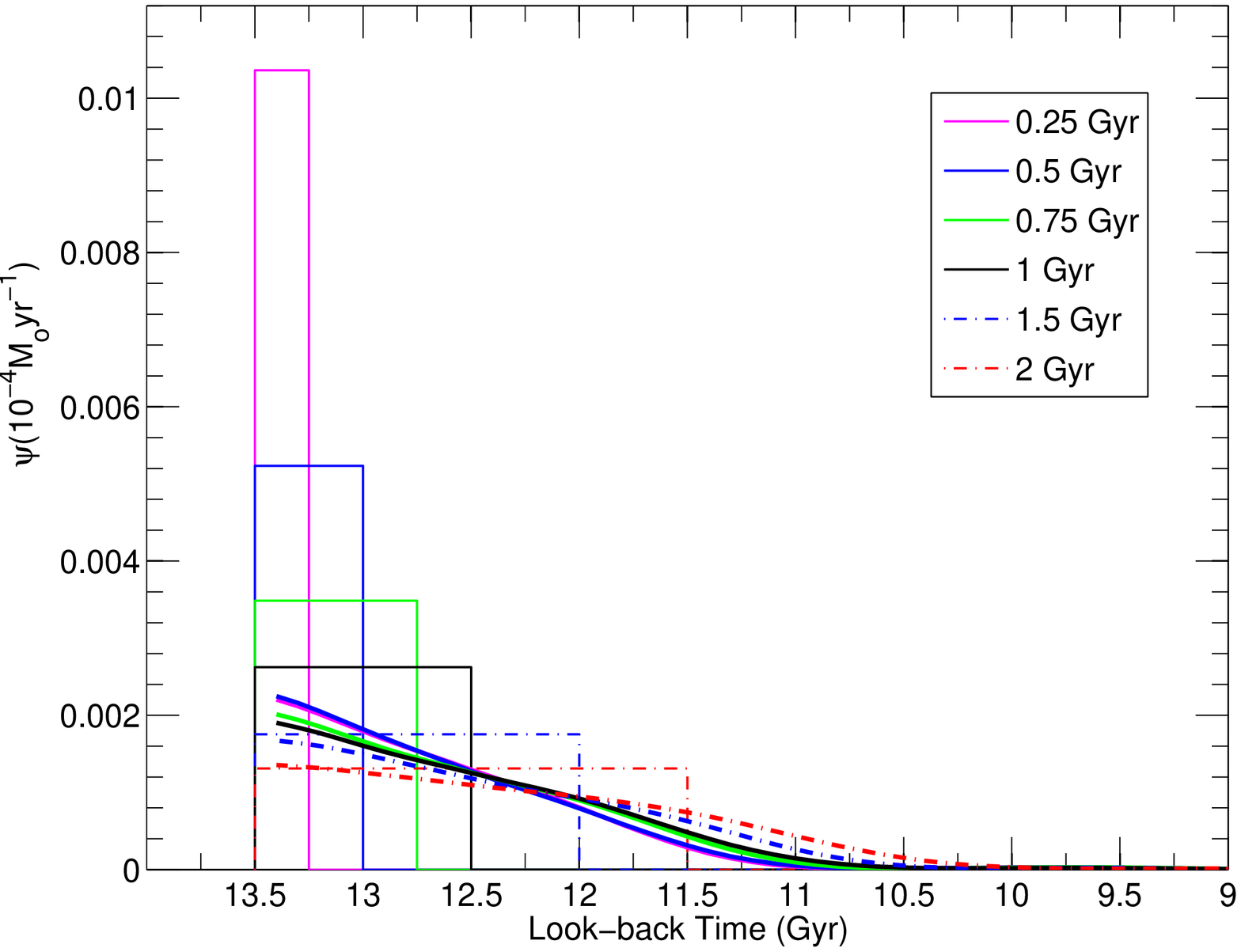}
   \caption{Recovering of an input simulated star formation burst. Input mock bursts with constant star formation rate and fixed time duration are shown (rectangular shape) with the recovered solution for each of them (smoothed curves).}
\label{fig:mocksfh}
\end{figure}

Observational effects on these mock bursts have been simulated using the completeness tests results described in \S~\ref{sec:case}. Then, the SFH of each mock population has been derived with the same procedure used for real data (i.e. the same \textit{bundles} and parameters). 
In order to quantify the uncertainties on the recovered $FWHM_{rec}$ we have repeated five times the simulation for each synthetic burst using different random number seeds in the IAC-star input parameters.
The mean recovered $FWHM_{rec}$ of each mock burst are: $2.5$, $2.6$, $2.8$, $2.9$, $3.2$, $3.3$ Gyr.
In Fig.~\ref{fig:mocksfh} are plotted both input mock bursts and the recovered SFHs associated to them. Since we have simulated an input constant star formation rate in a fixed time interval, input mock bursts have a rectangular shape, while, due to simulated photometric errors, the recovered SFHs result smoothed and broadened over a larger time interval.
Fitting a Gaussian profile to the Sextans observed $\psi(t)$ in the age range $9.5-13.5$ Gyr we estimate $\sigma=1.2$ Gyr, which  corresponds to a $FWHM_{obs}=2.8$ Gyr. 

\begin{figure}

\includegraphics[width=\columnwidth]{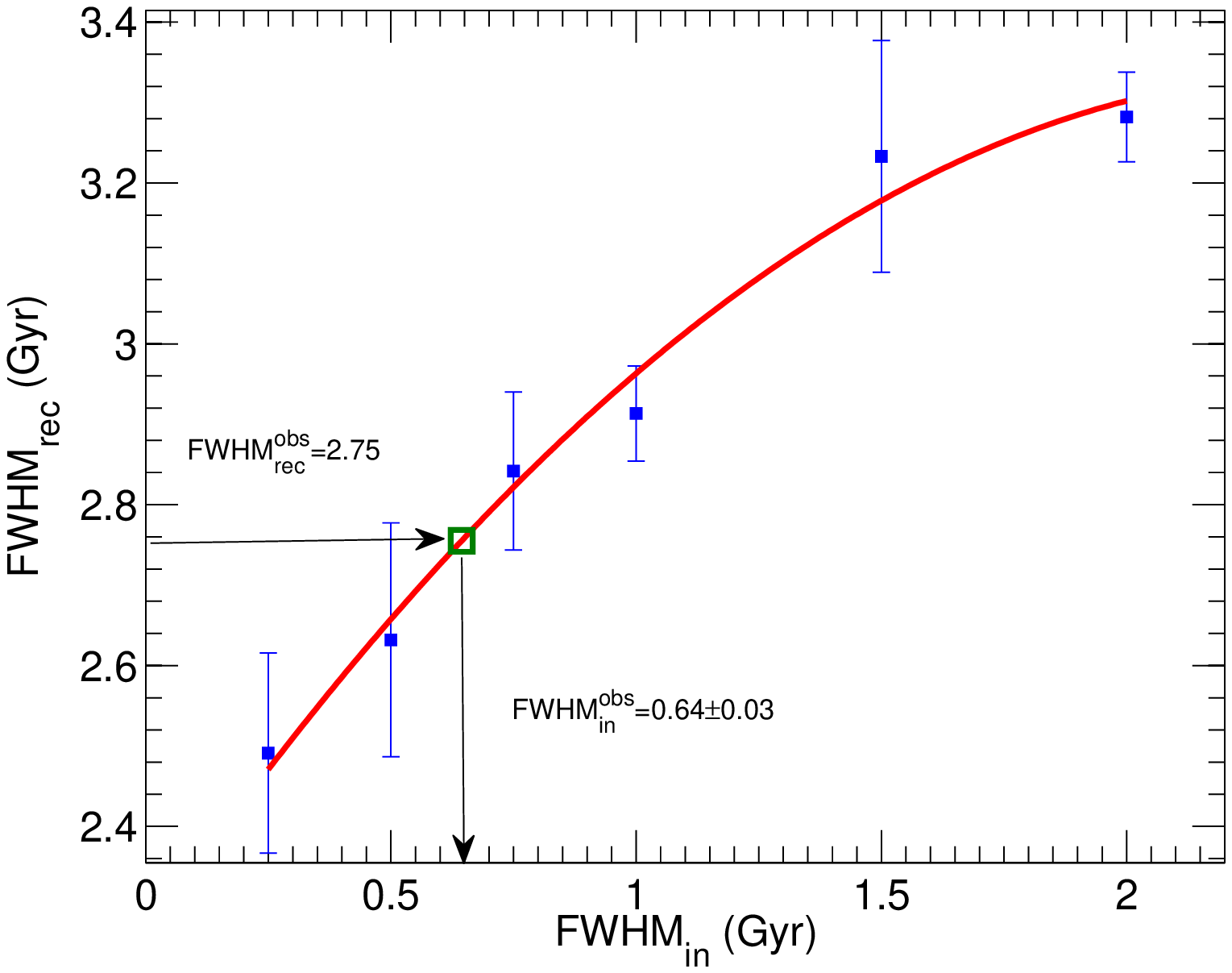}
   \caption{$FWHM_{in}$ is the Full Width at Half Maximum of the input mock bursts, $FWHM_{rec}$, the recuperated FWHM relative to the SFH of each mock bursts. Points have been fitted with a quadratic polynomial. Knowing the FWHM of the best SFH solution, making use of the intercept on the fitting red line (see green box) we have been able to confine the first star formation burst to a value of $FWHM_{in}^{obs}\sim0.64$ Gyr}
\label{fig:limits}
\end{figure}

In Fig.~\ref{fig:limits} we plotted the $FWHM_{in}$ of the mock bursts and their associated mean recovered $FWHM_{rec}$. Fitting the resulting data with a quadratic polynomial we are able to constrain the first star formation event to an hypothetical mock burst having a $FWHM_{in}=0.64\pm0.03$ Gyr.
The obtained result is consistent with the estimate of about $0.8$ Gyr obtained by \citet{2011ApJ...727...79K} via chemical evolution models.

Seen in this light, the star formation ended close to $12.9$ Gyr ago, while the end of the epoch of the reionization is fixed at $\sim12.77$ Gyr \citep{2001AJ....122.2850B}. 
This means that the star formation in Sextans was completed together to the end of the epoch of reionization.

\subsection{Radial SFH}
\label{sec:radial}
\begin{figure}

\includegraphics[width=\columnwidth]{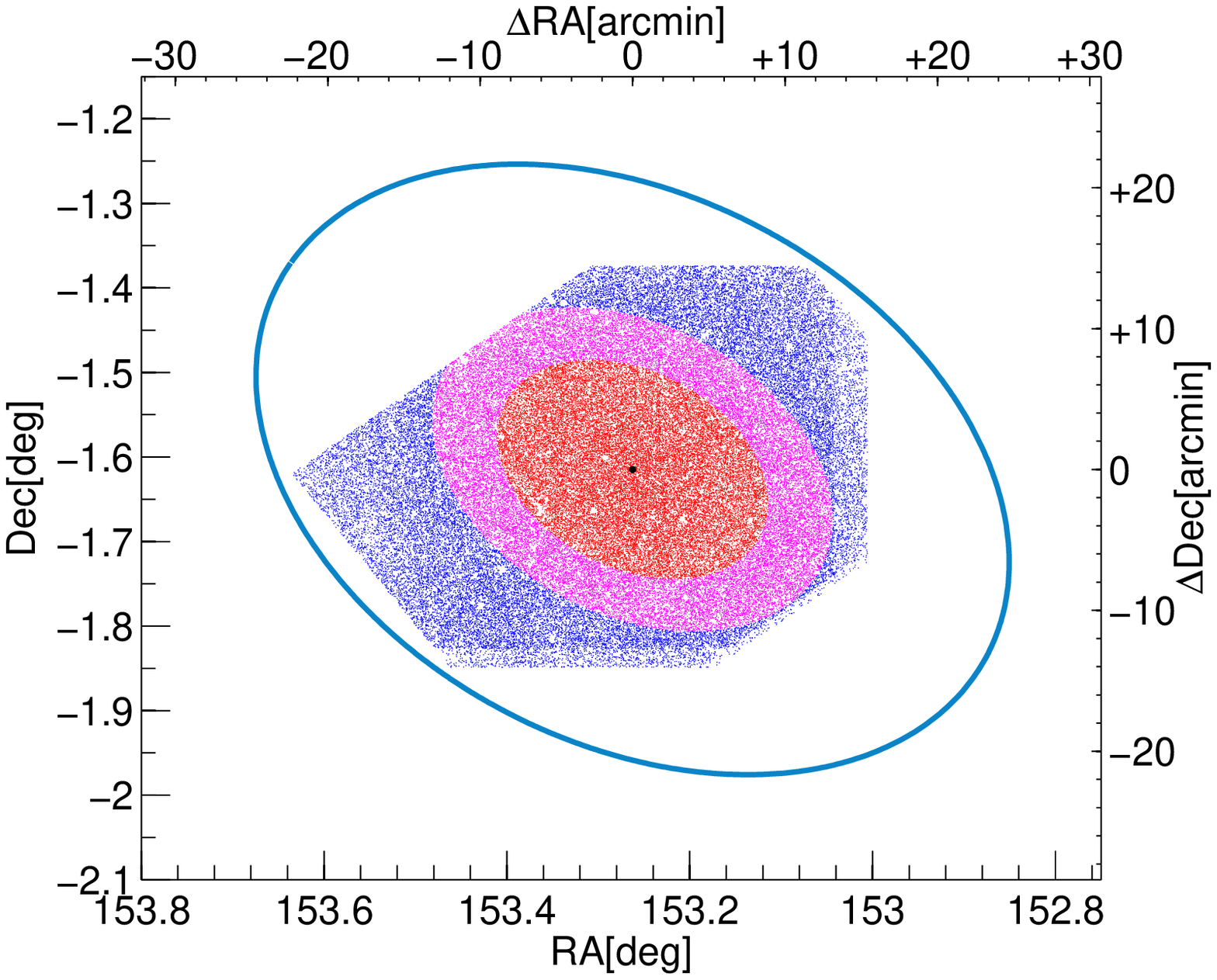}
   \caption{ Stars spatial distribution, the ellipse major axis are in the intervals $a\leq9.65$ arcmin for red points; $9.65<a\leq14.24$ arcmin for magenta points; and $14.24<a\leq25.86$ arcmin for blue points. The ellipse delimiting the core radius is plotted in light blue.}
\label{fig:spaz}
\end{figure}

\begin{figure}

\includegraphics[width=\columnwidth]{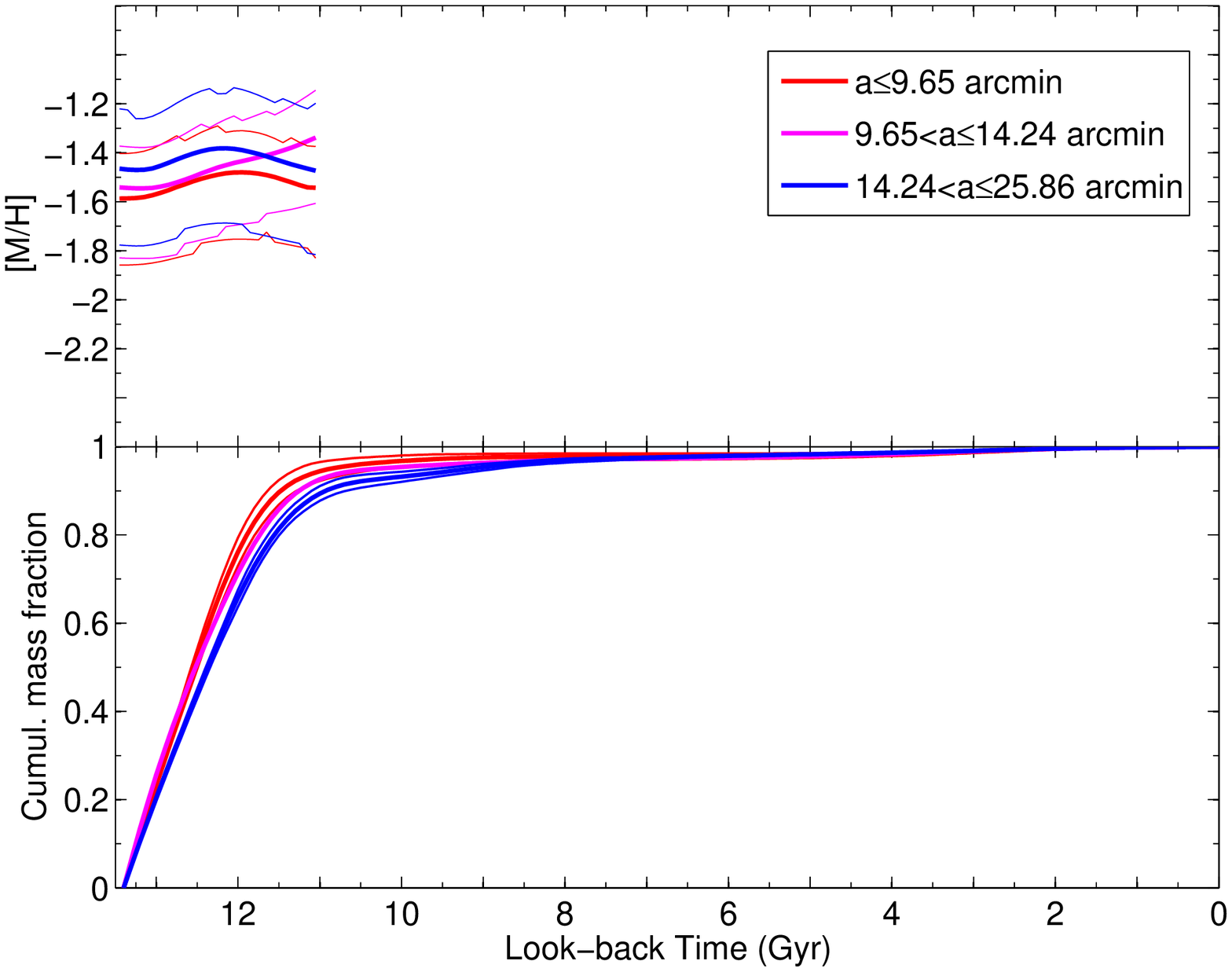}
   \caption{Upper panel: metallicity of the system as a function of the time for the three regions selected in Fig.~\ref{fig:spaz}. Lower panel: cumulative mass fraction as a function of the time. The same colour code as in Fig.~\ref{fig:spaz} has been used. For clarity only the error bands relative to innermost and outermost regions have been shown.}
\label{fig:mettot}
\end{figure}

In passing we made the attempt to investigate the radial variation of the SFH. In order to do this we divided the area into 3 elliptical regions, with delimiting major axis of $9.65$, $14.24$ and $25.86$ arcmin, see Fig.~\ref{fig:spaz}. The center assumed for the analysis is coincident with the one tabulated by \citet{1990MNRAS.244P..16I} in J2000.0 coordinates (153.2623,-1.6146). The position angle adopted is $\theta=56.7$ \citep{2016MNRAS.460...30R}.
The major axes have been fixed in order to have $\sim29000$ stars in each elliptical region for statistical consistency. For each region it has been derived the SFH using in each case the corresponding oCMD and as model CMD the one adopted for the total SFH.
In Fig.~\ref{fig:mettot} is shown the resulting metallicity as a function of time for all the $3$ regions. Within errors it is not detectable any metallicity gradient. The same can be visualized from the comparison of the cumulative mass fractions, thus indicating the absence of a radial variation of SFH. (lower panel Fig.~\ref{fig:mettot}).

\section{Discussion}
\label{sec:disc}
The resulting SFH is globally in agreement with the findings by \citet{2009ApJ...703..692L} who derived the SFH from the photometry presented in \citet{2003AJ....126.2840L}.
This is also confirmed by the mean metallicity we derived, equal to $[Fe/H]=-1.60\pm0.25$, which is in good agreement with the one estimated by \citet{2009ApJ...703..692L} in the central region of their field. 
\citet{2009ApJ...703..692L} measured that the metallicity of the stars is $[Fe/H]=-1.6$ in the central region, decreasing down to $[Fe/H]=-1.8$ in the outer region within the first Gyr.
From our analysis, instead, emerges that metallicity in the inner core radius of Sextans does not present a radial gradient, as shown in Fig.~\ref{fig:mettot}, though our data span more or less the same radius of \citet{2009ApJ...703..692L}. 
Moreover, the metallicity gradient measured between the innermost and the more external region in \citet{2009ApJ...703..692L} is of the order of $\sim0.2$ dex, a value close to the error in determination of our metallicity.
This difference in the detection of the gradient could be related to a poor statistics in the 
number of stars used by \citet{2009ApJ...703..692L} for the calculation of the SFH in each region, of the order of $\sim5000$ stars, while we adopted $\sim30000$ stars for each of our three regions.
This could be also related to photometric data not deep enough, so that, the SFH is affected by large errors. In fact, in that work, the completeness for magnitude $I\sim23.5$ is at $\sim55\%$, while in the present work, the completness at these magnitudes is more than $95\%$. This strongly influence the result and the associated error. 
Another important aspect is that \citet{2009ApJ...703..692L} assumed that Sextans started forming stars $15$ Gyr ago, while we constrained the age of the oldest acceptable population to $13.5$ Gyr. 
 
Our results on metallicity are supported by the fact that Sextans presents almost a flat $[Fe/H]$ radial distribution from the spectroscopic measurements conducted by \citet{2011ApJ...727...78K} over an area very similar to the one sampled in this work. The authors concluded that the lack of gradients, also seen in Canes Venatici I and Ursa Minor, indicates a star formation event significantly shorter than $1$ Gyr. By means of chemical evolution models, they find also for Sextans, a duration of the star formation of just $0.8$ Gyr, hence in striking agreement with what we found in our analysis.

\begin{figure}

\includegraphics[width=\columnwidth]{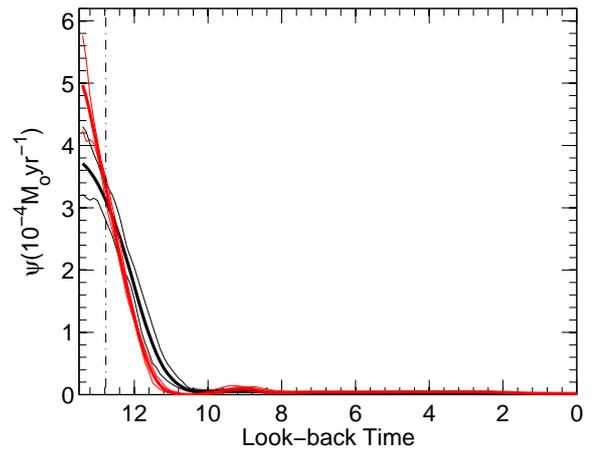}
   \caption{Original SFH, obtained without assuming any a priori MDF (in black) and the SFH derived imposing the MDF by \citet{2011ApJ...727...78K} (in red).}
\label{fig:sfhcomparate}
\end{figure}

\begin{figure}

\includegraphics[width=\columnwidth]{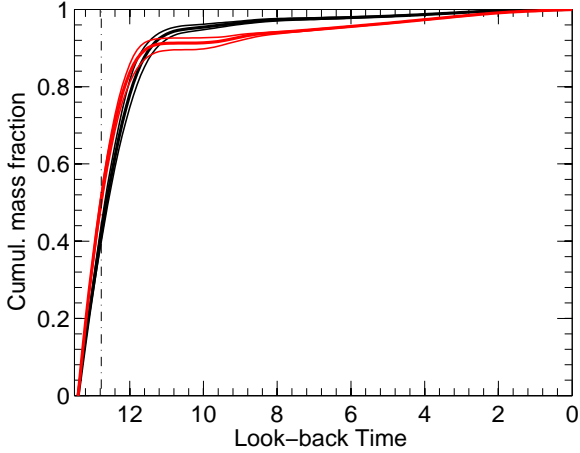}
   \caption{Cumulative mass fraction as a function of the lookback time, color coded as in Fig.~\ref{fig:sfhcomparate}.}
\label{fig:cumulata}
\end{figure}

But the metallicity distribution found by \citet{2011ApJ...727...78K} peaks at $\sim-1.8$, $0.2$ dex lower than  our mean metallicity determination. This difference is consistent with the metallicity bin we used in the derivation of the SFH, which is $\sim0.0002$.
As a further test, to see if the SFH is affected by this metallicity difference, we rederive it imposing an input model whose stars follow the MDF by \citet{2011ApJ...727...78K}. 
In Fig.~\ref{fig:sfhcomparate} are shown the original SFH, obtained without assuming any a priori MDF (in black) and the SFH derived imposing the MDF by \citet{2011ApJ...727...78K} (in red). In Fig.~\ref{fig:cumulata} the cumulative mass fraction as a function of the lookback time are also shown, color coded as in Fig.~\ref{fig:sfhcomparate}. The SFH in both cases is very similar in shape, demonstrating that assuming a MDF as in \citet{2011ApJ...727...78K} does not change the results of the present work. Moreover, assuming this MDF the star formation appears even steeper than the original one, thus confirming the trend already found without assuming any MDF for the input model.
It is worth noting that even a mean metallicity of $-1.6$ for Sextans would be consistent with the trend in the metallicity-luminosity plane for galactic dSph (see \citet{2003AJ....125.1926G} and \citet{2012AJ....144....4M}).



The question arising now is why, as detected with larger surveys, does Sextans present clear signs of the presence of a younger (and more metal rich) stellar population toward its center, that translates in an age and metallicity gradient. (see \citealt{2001AJ....122.3092H}, \citealt{2001MNRAS.327L..15B}, \citealt{2002ASPC..274..421P}, \citealt{2003AJ....126.2840L}, \citealt{2004MmSAI..75..110R}, \citealt{2017MNRAS.tmp...97O})

\citet{2017MNRAS.tmp...97O} through the analysis of the various evolutionary phases along a radius more extended than the tidal radius, have been able to detect an age gradient. Relatively younger stars ($\sim 10$ Gyrs) are more centrally concentrated than old stars ($\sim 13$ Gyrs). Through our detailed analysis, limited to the core radius, we have not detected any stellar populations gradient, so we conclude that the gradient found by \citet{2017MNRAS.tmp...97O} has its origin after the core radius.

This point has been discussed also for the case of the isolated dSph Tucana \citep{2010ApJ...722.1864M}, that resemble for some aspects the case of Sextans. Tucana SFH is characterized by a unique and pronounced event of star formation and a steep rising of metallicity in correspondance of the burst followed by almost no evolution.
But hints of the presence of two generations of stars have been found in the complex morphology of the HB in Tucana \citep{2010ApJ...722.1864M}.
It is reasonable that the second generation formed towards the center, since newly processed gas would concentrate there.

This idea is supported by the evidences found in the MDF by \citet{2011ApJ...727...78K}. 
This presents, in fact, a small bump at $[Fe/H]\sim-3$, so at extremely low metallicities. This bump is interpreted by the authors as the sign of a rapid burst happened at early times followed by an epoch of minimal star formation, maybe due to SNe type II strong ejecta.

\citet{2011MNRAS.411.1013B} found that within $0^{\circ}.8$ from the center of Sextans, there is an high metalicity spread, with $-3.8<[Fe/H]<-1.4$ and an average $[Fe/H]\sim-1.9$, so partially in agreement with our results, even though our mean metallicity is $\sim0.3$ dex higher. In the outer region, ($R>0^{\circ}.8$), the metal poor population is predominant. Evidently, these stars underwent a very quick formation, while the stars in the inner part a more prolonged one.

This picture would suggest a first brief event of star formation which produced the diffuse large scale metal poor population and a more extended star formation that generated the centrally concentred metal rich population.
From the study performed by \citet{2013ApJ...778..103H} on the radial variation of SFH of dwarf galaxies emerges that Sextans matches the case of a typical dSph characterized by an extended old stellar population of $\sim 13$ Gyr at all galactocentric radii and a very small age gradient towards the center. 

In the works by \citet{2005ApJ...629..259R} and \citet{2011ApJ...741...18B} it is resumed a very stimulating aspect: the early and single star formation burst of Sextans could be an array of short bursts that we are not able to resolve with the actual techniques.
The authors conclude that the metal spread and abundances, typical of dSph of  the MW, are not necessarily produced by a prolonged and continous star formation of more than $2-4$ Gyrs (see e.g. \citet{2004ApJ...610L..89G}); similar values can be obtained in the regime of bursty and multiple events of star formation with timescales down to $50$ Myrs \citep{2001ApJ...558..351M}.
This is strictly connected to the iron enrichment by SNe Ia that is uncertain and dependant on the mode of star formation.
\citet{2001ApJ...558..351M} estimated that this timescale varies from 40 to 50 Myr for an instantaneous starburst to 0.3 Gyr for a typical elliptical galaxy, thus far from our time resolution.

Sextans stopped early to form stars and it is of great interest to investigate if this phenomenon is linked in any way to the ultraviolet (UV) cosmic reionization, or if it is mainly due to local causes, such as intense galactic winds and SNe feedback (\citealt{2011ApJ...730...14H}, \citealt{2016ApJ...823....9A}, \citealt{2004MNRAS.351.1338L}).
 
When considering the results of the retrieved SFH as well as data in literature, the question arising is whether Sextans is 'true fossil' of the pre-reionization era in the sense introduced in \citet{2005ApJ...629..259R}: a dwarf that has experienced more than the $70\%$ of its  star formation before the end of the reionization and that has a luminosity $L_V<10^6 \,L_\odot$.
Sextans satisfies the second condition, since it has a luminosity $L_V=(4.1\pm1.9)\times 10^5 \,L_\odot$ \citep{1995MNRAS.277.1354I}.  
Regarding the first condition, the galaxy formed $70\%$ of its stellar mass $\sim12$ Gyr ago as derived from the SFH, but, as discussed in \S~\ref{sec:confining}, we have been able to confine the initial event of star formation to $\sim0.6$ Gyr, so, the above mass fraction would be completed surely within the reionization epoch.

From these two facts Sextans can really be classified a \textit{true fossil} of the pre-reionization era.
Moreover, the work by \citet{2005ApJ...629..259R} corroborates the hypothesis according to which internal feedback, such as photoheating by the stars inside the galaxy and SNe explosions, have been the principal causes of suppression of star formation, since true fossils are the result of feedback processes in action prior to reionization. So that, they are expected to evolve passively after the end of the reionization era because of the exaustion of gas; this is what the results of our analysis of Sextans are suggesting.

In order to investigate this hypothesis we have calculated, as outlined in \citet{2011ApJ...730...14H}, the mechanical luminosity of the SNe released during the main star formation episode. This quantity together with the value of the mass of gas of the galaxy can be compared with the results presented in \citet{1999ApJ...513..142M} to discern the modality according to which Sextans had lost mass in the past.
Using the results from the SFH, we have calculated that a total of  $9.2\times10^{5}\,{\rm M}_{\odot}$ of gas were converted into stars. We have derived this value by scaling the SFH obtained from our observations to the whole galaxy using the King profile as obtained in \citet{2016MNRAS.460...30R}.
We obtained a total of $1.6\times10^{4}$ SNe assuming a minimum progenitor mass for core collapse SNe of $6.5\,{\rm M}_{\odot}$ \citep{2005essp.book.....S}, and $8.7\times10^{3}$ SNe in the case of a progenitor of $10\,{\rm M}_{\odot}$. 
Assuming an energy release per SN of $10^{51}$erg \citep{1999ApJS..123....3L} and a duration of the episode of $0.6$ Gyr, we have calculated a total mechanical luminosity released during the old episode of $L_{w}=8.2\times10^{38}$ erg/s and $L_{w}=4.6\times10^{38}$ erg/s, respectively for $6.5$ and $10\,{\rm M}_{\odot}$. 
From the comparison of the above values with the model results of \citet{1999ApJ...513..142M} shown in their Figure 1, it is clear that Sextans places in the region of blow-away regime in concomitance to the blow-out/mass loss regime. 
This result indicates that effectively SNe could have played an important role in the fate of Sextans by removing almost completely the gas component, so preventing a star formation extended over time (\citealt{1986ApJ...303...39D},\citealt{2001ApJ...558..598K}, \citealt{2002MNRAS.335..335C}). From our analysis we can not quantify how much reionization has influenced the SFH of Sextans dSph, even though it has persisted during all the early evolution of the galaxy. From the simulations of \citet{2010MNRAS.402.1599S} it is not expected that Sextans could be able to form stars after the epoch of reionization due to its relatively low mass. Moreover, there are also external effects such as the ram pressure stripping \citep{2013MNRAS.433.2749G} that have to be taken into account. Concerning the fact that Sextans' SFH could have been influenced by tidal interactions, from structural studies such as the one by \citet{2016MNRAS.460...30R} there are no clear signs of an undergoing tidal disruption from the MW. 


\section{Summary and Conclusions}
\label{sec:conc}
We derived the star formation history (SFH) of the Sextans dwarf spheroidal galaxy based on deep archive $B$,$I$ photometry taken with Suprime-Cam at Subaru telescope. The data are limited to the core radius of the galaxy. We have not detected any metallicity gradient along the considered radial distance interval within the errors of our SFH. 
We have been able to constrain the duration of the main burst of star formation to $\sim0.6$ Gyr, thus indicating that the Sextans dwarf spheroidal stopped forming stars $\sim12.9$ Gyr ago before the end of the reionization epoch.
From our analysis based on the model results by \citet{1999ApJ...513..142M} and the calculation of the mechanical luminosity released from SNe during the brief episode of star formation, we can advance the hypothesis that Sextans run out most of its gas reservoirs before the end of the reionization due to gas outflows induced by SNe of type II, capable to throw out of the dwarf potential well the newly produced enriched elements and the remaining gas.

\section*{Acknowledgements}
The authors thank the anonymous referee for the constructive comments that improved the quality of the present manuscript.
MB, SLH, SC, AA and GP  acknowledge support from the Spanish Ministry of Economy and Competitiveness (MINECO) under grant AYA2013-42781.
Based on data collected at Subaru Telescope and obtained from the SMOKA, which is operated by the Astronomy Data Center, National Astronomical Observatory of Japan.
This research used the facilities of the Canadian Astronomy Data Centre operated by the National Research Council of Canada with the support of the Canadian Space Agency. 
This research has made use of the NASA/IPAC Extragalactic Database (NED) which is operated by the Jet Propulsion Laboratory, California Institute of Technology, under contract with the National Aeronautics and Space Administration.









\bsp	
\label{lastpage}
\end{document}